\documentclass[conference]{IEEEtran}

\usepackage[pdftex]{graphicx}

\usepackage{amsmath,amsthm,amsbsy,amscd,mathrsfs}
\usepackage{amssymb}
\usepackage{graphicx,color,verbatim}
\usepackage{mathtools}
\usepackage{booktabs}
\usepackage{cite}
\usepackage{multirow}

\theoremstyle{plain}

\newtheorem{lemma}{Lemma}
\newtheorem{proposition}{Proposition}

\theoremstyle{remark}
\newtheorem{remark}{Remark}
\theoremstyle{definition}
\newtheorem{defn}{Definition}
\newtheorem{example}{Example}

\def\bb{\mathbb}

\def\ZZ{\bb{Z}}
\def\NN{\bb{N}}

\def\QQ{\bb{Q}}
\def\RR{\bb{R}}
\def\NN{\bb{N}}
\def\ZZ{\bb{Z}}

\newcommand{\LB}{\ensuremath{\Lambda_B}}
\newcommand{\LE}{\ensuremath{\Lambda_E}}

\DeclareMathOperator{\diag}{diag}

\DeclareMathOperator{\Prob}{P}
\DeclareMathOperator{\Exp}{E}

\DeclareMathOperator{\vol}{vol}
\DeclareMathOperator{\ECDP}{ECDP}

\DeclareMathOperator{\Short}{S}

\begin{document}
%
\title{Well-Rounded Lattices for Reliability and Security in Rayleigh Fading SISO Channels}

\author{\IEEEauthorblockN{Oliver Wilhelm Gnilke, Ha Thanh Nguyen Tran, Alex Karrila, Camilla Hollanti}
\IEEEauthorblockA{
	Department of Mathematics and Systems Analysis\\
	Aalto University School of Science, Finland\\
Emails: \{oliver.gnilke, ha.n.tran, alex.karrila, camilla.hollanti\}@aalto.fi}
}


%


\maketitle

\begin{abstract}
	For many wiretap channel models asymptotically optimal coding schemes are known, but less effort has been put into actual realizations of wiretap codes for practical parameters. 
	Bounds on the mutual information and error probability when using coset coding on a Rayleigh fading channel were recently established by Oggier and Belfiore, and the results in this paper build on their work. However, instead of using their ultimate inverse norm sum approximation, a more precise expression for the eavesdropper's probability of correct decision is used in order to determine a general class of good coset codes. The code constructions are based on well-rounded lattices arising from simple geometric criteria. In addition to new coset codes and simulation results, novel number-theoretic results on well-rounded ideal lattices are presented.
\end{abstract}


%
\IEEEpeerreviewmaketitle

\section{Introduction}
In the wiretap setting it is assumed that the same message is transmitted over two different channels, a channel to an intended/legitimate receiver Bob and a different channel to an eavesdropper Eve. Three contradicting objectives are simultaneously tried to be achieved:  A high information rate between the sender and Bob, high reliability at the legitimate receiver, and minimal mutual information between the message and the output at the eavesdropper.

Several different design criteria have been derived for secure SISO wiretap channels, such as an eavesdropper's probability bound and an inverse norm sum approximation \cite{BelfioreOggierRayleigh}, and an information bound \cite{Mirghasemi,ling,luzzi, AlexFlat}, showing that both probability and information are bound by the flatness factor. 
The current design criteria for wiretap channels, such as the inverse norm sum, are based on loose approximations and are not very reliable in terms of comparing different codes. Hence, it is desirable to derive new design criteria from tighter approximations. Motivated by this, we study the eavesdropper's probability of correct decision and conclude that the coset codes used for the transmissions should arise from \emph{well-rounded} (WR) lattices. By using WR sublattices of rotated $\ZZ^n$ lattices, one can simultaneously try to optimize for both low and high signal-to-noise ratio (SNR). Furthermore, this allows for the use of a skewed lattice for Eve while maintaining an orthogonal lattice structure for Bob, as suggested in \cite{AlexSkew}.
\subsection{Related Work and Contributions}
In this paper we reexamine design criteria suggested in other works \cite{BelfioreOggierRayleigh, Mirghasemi}. We propose a new, simpler and more geometric criterion, based on well-rounded (WR) lattices. The construction of WR lattices by algebraic means is investigated and a result on WR principal ideal lattices is obtained. After constructing several examples in practical dimensions, their superiority is supported by several simulations.
\section{Preliminaries}
\subsection{Wiretap Channel}
We consider a wiretap channel as described by Wyner \cite{Wyner}. A sender, Alice, transmits data over a possibly noisy channel to a receiver Bob and a second noisy channel to an eavesdropper Eve exists. The common assumption made is that the channel to the eavesdropper has lower SNR, or is in some other way more degraded, than the channel to the legitimate receiver. Wyner investigated the possibility of transmitting data to the receiver while having negligible mutual information with the eavesdropper. He could show that a non-zero secrecy capacity exists when a binary symmetric channel is considered. The mutual information between variables $X$ and $Y$ is defined in the usual way as 
\begin{equation} I(X,Y):= \sum_X \sum_Y \Prob[X,Y] \log\left( \frac{\Prob[X,Y]}{\Prob[X]\Prob[Y]} \right), \end{equation}
see \cite{InfoSec} for more details.
We will consider a single-input single-output (SISO) fast Rayleigh fading channel model \cite{Viterbo_green}, where the information is mapped to vectors in a codebook $x \in \mathcal{C} \subset \RR^n$ that are then component-wise sent through the channel. The vectors received by Bob and Eve are respectively given by
\begin{equation} 
y=H_B x + e_B  \text{ and } z=H_E x + e_E
\end{equation}
where $H_*=\diag(h^*_i)$ is a $n \times n$ diagonal matrix with $h^*_i$ being Rayleigh distributed fading coefficients with second raw moment $\Exp[(h^*_i)^2]=1$ and $e \in \RR^n$ a Gaussian distributed error vector where each entry has variance $\sigma_*^2$. 


It is commonly assumed that Bob, by virtue of his superior channel, is able to decode correctly with high probability. All elements relevant to the security of a scheme are related to the channel to Eve and we will therefore from now on only consider it and suppress the index $E$.
Even though Eve's low SNR might lead her to decode incorrectly she might still gather information from the transmission, as highlighted in the following example.

\begin{example} \label{ex:border}
	Given the one dimensional example where $\mathcal{C}=\{-2,-1,0,1,2\}$, consider a channel where $H=1$ and $x=1$ was sent but Eve receives a vector closer to $2$ due to the error term $e$. She decodes to $2$, which is of course incorrect, but the mutual information is far from being negligible, since she still learned that the sent vector is highly unlikely to have been $-2, -1$ or $0$.
\end{example}

In a perfect setup Eve would not be able to gather any information from the vector $z$. Having zero mutual information is equivalent to every codeword being equally likely to have been sent. Therefore a different strategy, coset coding, is employed where one message $m$ is represented by several different codewords $[m] \subset \mathcal{C}$ in the codebook $\mathcal{C}$. The probability that $m$ has been sent is given as the sum of the probabilities of the different codewords that represent that message. This strategy of course trades data rate for security. Information theoretically we try to increase the entropy for the random variable $z$ for fixed $x$.

\section{New Design Criteria for Nested Lattice Coset Codes}
To achieve secret transmission of information, coset coding in nested lattices has been suggested by Oggier and Belfiore \cite{OggierCoset}, based on ideas in \cite{Wyner} and \cite{Wyner2}. We begin by introducing lattices and some necessary notation.

\begin{defn}
	A \emph{lattice} $\Lambda$ is the $\ZZ$-linear span of a set $\{b_1, \dots, b_m \}\subset \RR^n$ of linearly independent basis vectors. We call $\Lambda$ a \textit{full-rank} lattice if $m=n$.
\end{defn}


The squared length of a shortest (non-zero) vector in $\Lambda$, the \textit{minimum distance}, is denoted by $\lambda_1(\Lambda)$. 
Each $x \in \Lambda $ such that $\|x\|^2 =\lambda_1(\Lambda)$  is called a \textit{minimal} vector, and the set of minimal vectors of $\Lambda$ is denoted by $\Short(\Lambda)$. The volume of a lattice $\vol(\Lambda):=|\det((b_i)_i)|$ where $\{b_i\}_i$ is any basis for $\Lambda$ is an important invariant.
The Hermite constant in dimension $n$ is defined as
\begin{equation} \gamma_n:=\max_{\Lambda \subset \RR^n} \frac{\lambda_1(\Lambda)}{\vol(\Lambda)^\frac{2}{n}}.\end{equation}

\subsection{Nested Lattice Coset Coding}
In nested lattice coset coding the codebook consists of vectors from a lattice $\Lambda_B$. 
This lattice is chosen such that Bob is able to decode correctly with high probability. 
A second lattice $\Lambda_E \subset \Lambda_B$ is then chosen and every possible message is mapped to an element in the quotient $\Lambda_B / \Lambda_E$. Consequently, the information rate is determined by the index $[\Lambda_B : \Lambda_E]:=\frac{\vol(\LE)}{\vol(\LB)}$, not the actual codebook size. The vector $x$ is chosen as a random representative of the coset belonging to the intended message. Hence, a codeword $x$ is the sum of the message $m \in \Lambda_B / \Lambda_E$ (for a fixed shortest set of representatives) and a random vector $r \in  \Lambda_E$. In practice we restrict $x$ to a finite region. This is done by restricting the coefficients in the linear combinations of the basis elements to a finite signaling set $S \subsetneq \mathbb{Z}$.

For a finite codebook $\mathcal{C} \subsetneq \RR^n$ the data rate $R:=\frac{1}{n}\log_2(|\mathcal{C}|)$ in bits per channel use (bpcu) is split between the information rate $R_i$ from Alice to Bob, \emph{i.e.}, the actual amount of information transmitted, and random bits $R_c$ added to confuse the eavesdropper
\begin{equation} R=R_i+R_c= \frac{1}{n}\log_2([\LB: \LE]) + \frac{1}{n}\log_2\left( \frac{|\mathcal{C}|}{[\Lambda_B : \Lambda_E]}\right). \end{equation}
Increasing the number of coset representatives and thus increasing $R_c$ reduces border effects but increases the average energy by increasing the codebook size.
\subsection{Correct Decoding Probability}
In several papers \cite{luzzi, ling} a connection between Eve's correct decoding probability (ECDP) and the mutual information has been established. Thus, we can use Eve's correct decoding probability $\ECDP$ as a measure of how much information she can glean from $z$. 

We point out that even if the probability would be quite high, the mutual information can still be zero, meaning Eve gains no information even though occasionally decoding correctly. Notice also that due to coset coding, there will be a coset representative close-by regardless of how big the noise is, as demonstrated by the lower bound $\frac{1}{[\Lambda_B : \Lambda_E]}$. This is where coset codes crucially differ from traditional lattice codes.

An analytic approximation for the correct decoding probability is developed in \cite{BelfioreOggierRayleigh} as  
\begin{small}
	\begin{equation} \label{analytic} \ECDP=\left( 2\sigma_E \right)^{-n} \vol(\Lambda_B) \sum_{r \in \Lambda_E} \prod_{i=1}^{n} \left( 1+ \left(\frac{r_i}{\sigma_E}\right)^{2} \right)^{-\frac{3}{2}} .
	\end{equation}
\end{small}
Here $n$ is the dimension of the lattices involved. The authors perform further approximations to come up with the so-called \emph{inverse norm sum}, also investigated in, \emph{e.g.}, \cite{Karpuk2014,Roope}. Here, we do not take these further  steps, but will analyze a more precise version of the probability expression as explained below.

Following the idea of $i$-th coding gains in \cite{ithgain} we expand the product and bound from below by ignoring everything but the constant, the linear and the leading term
\begin{small}
	\begin{equation} \label{eq:estimate} \prod_{i=1}^{n} \left( 1+ \left(\frac{r_i}{\sigma_E}\right)^{2} \right)^{-\frac{3}{2}} \hspace{-10pt} \leq \left( 1 + \sum_{i=1}^{n} \left(\frac{r_i}{\sigma_E}\right)^{2} +  \prod_{i=1}^{n} \left(\frac{r_i}{\sigma_E}\right)^{2} \right)^{-\frac{3}{2}}. \end{equation}
\end{small}
The linear term corresponds to the squared length of the vector $\sigma_E^{-1} r$, while the leading term is given by the squared product distance $d_{p,\min}(v):=\prod_{i=1}^{n}|v_i|$ of $v=\sigma_E^{-1} r$. Rotations of $\ZZ^n$ that maximize the minimum product distance have been investigated in several publications such as \cite{Viterbo_green} or \cite{DaveRotate}.
These increase reliability in channels with high SNR, or low $\sigma_E$, where the leading term is dominant. In channels with lower SNR the linear term becomes more prominent and we should choose our lattice $\LE$ such that it is maximized.

Since the shortest vectors of $\LE$ contribute the most by the estimate (\ref{eq:estimate}) it is beneficial to maximize their length. This is equivalent to increasing the sphere packing radius of the lattice, \emph{i.e.}, maximizing the minimal length of its vectors. 

Although optimal sphere packing lattices would provide the longest minimal vectors achieving the Hermite constant, these are not always available or even the best choice. 
Even when an optimal integer lattice exists, \emph{e.g.}, $D_4$ or $E_8$ with suitable scaling, they only provide us with limited choices for indices. We therefore suggest a more general class of lattices, namely well-rounded lattices, which are available for many different parameters. In this paper we will focus on (possibly rotated) sublattices of $\ZZ^n$. The rotations are algebraic, so the lattices remain integral and can be used to guarantee good performance in the high SNR regime. We use the standard pulse amplitude modulation (PAM) for the lattice coordinates 
We sum up our design criterion in the following proposition, antedating the definition in the following section, but postpone a rigorous proof to an extended journal version.
\begin{proposition}
	Well-rounded lattices optimize expression \eqref{eq:estimate} in the low SNR regime and give rise to good sublattices for coset coding.
\end{proposition}

\section{Well-Rounded Lattices}

\begin{defn}
	A lattice $\Lambda$ is called \textit{well-rounded} (abbreviated WR) if the set $S(\Lambda)$ of minimal vectors contains $n$ linearly independent vectors. 
\end{defn}

The set of minimal vectors $S(\Lambda)$ does not necessarily form a basis for $\Lambda$ {\cite[Chapter 2]{PhongNguyen}}. They are known to form a basis for all $n \le 4$ as mentioned in \cite{McMullenMinkowski}. This motivates the following stronger definition.
\begin{defn}\label{def2}
	A lattice $\Lambda$ is called \textit{(strongly) well-rounded} if the set of minimal vectors $S(\Lambda)$ generates $\Lambda$. 
\end{defn}

If a lattice is WR, then the set of minimal vectors $S(\Lambda)$ generates a sublattice of $\Lambda$ that is also WR. Hence, from now on, we only work with WR lattices as in Definition \ref{def2}.

More generally, WR lattices are of interest in investigations of sphere packing, sphere covering, and kissing number problems as well as in coding theory as shown in \cite{McMullenMinkowski,Blichfeldt,Martinet-Jacques}. A particularly interesting class of WR lattices are the integral well-rounded (IWR) lattices. The properties of WR lattices in the plane which come from ideals in quadratic fields have been studied by Fukshansky \emph{et al.} 
In \cite{FukIdeal2} and \cite{FukIdeal}, the authors presented a characterization of WR ideal lattices in the plane and proved that even asymptotically a positive proportion of real and imaginary quadratic number fields contain ideals giving rise to WR lattices. We provide two new related results in the next section. 

There is an easy criterion that relates the volume of a WR sublattice to the length of its minimal vectors using Hermite constants. 
\begin{lemma}\label{hermite}
	For a full rank WR lattice of volume $V$ it holds that $  V^{\frac{2}{n}} \leq \lambda_1 \leq \gamma_n  V^{\frac{2}{n}}   $, 
	where $\gamma_n$ is the Hermite constant for dimension $n$.
\end{lemma}


\subsection{Well-Rounded Ideal Lattices}
WR lattices that come from ideals of number fields are also called \textit{well-rounded ideal lattices}. Well known examples of these lattices are the ring of integers and its ideals of cyclotomic fields \cite{FukIdeal}. In \cite{FukIdeal2} and \cite{FukIdeal}, the authors proved the existence of infinitely many real and imaginary quadratic fields that contain WR ideal lattices and studied their properties. A sufficient (resp.  equivalent) condition for a positive square-free integer $D$ such that the quadratic field $\mathbb{Q}(\sqrt{D})$ (resp.  $\mathbb{Q}(\sqrt{-D})$) contains WR ideal lattices was also given. However, for an arbitrary number field of degree at least three, the existence and structure of WR ideal lattices are unknown. 

In this section, we further concentrate on real quadratic fields $F$. Regarding computational aspects, the norms of ideals are frequently considered in comparison with the discriminant of $F$. The relevance of the norm for our purposes arises from the fact that it corresponds to the nesting index and hence relates to the information rate, as we will see in this section.   Here, we first present a result saying that WR ideal lattices of $F$ have large norms compared to the discriminant. This is a nice result also from a practical point of view, since optimizing the minimum product distance is equivalent to minimizing the discriminant \cite{Viterbo_green}, whereas large norms are preferable due to their relation to the information rate. 

In addition, we discuss a class of WR lattices that are generated from \emph{principal} ideals. This also provides an answer to the question proposed in {\cite[Question 2]{FukIdeal2}} about the existence of WR principal ideal lattices of real quadratic fields. 

\begin{defn}\label{embb}
	Let $F$ be a totally real number field of degree $n$. There are exactly $n$ distinct field homomorphisms $\sigma_i: F \longrightarrow \mathbb{R}$ for $i=1, \ldots, n$. 
	The map
	$\sigma: F \longrightarrow \mathbb{R}^n$ 
	defined by $\sigma(\alpha) = (\sigma_1(\alpha), \ldots  \sigma_n(\alpha))$ is called the \textit{canonical embedding} of $F$. 
\end{defn}

Let $F$ be a totally real number field of degree $n$ with the ring of integers $O_F$. The images of ideals in $O_F$ under $\sigma$ are lattice in $\mathbb{R}^n$. We denote by $\Lambda_I = \sigma(I)$ the lattice in $\mathbb{R}^n$ corresponding to the ideal $I$ in $O_F$ and $\Lambda_F =\sigma(O_F)$. Then $\Lambda_I$ is a sublattice of $\Lambda_F$.  Moreover, the norm $N(I)$ of $I$ is also the index of $\Lambda_I$ in $\Lambda_F$ and hence it gives the information rate as described below.  

\begin{proposition}\label{norm}
	Let $\Lambda_E=\Lambda_I$ and $\Lambda_B = \Lambda_F$. Then $N(I)=[\Lambda_B:\Lambda_E]=2^{nR_i}$. 
\end{proposition}

\begin{remark}
	Under the canonical embedding, the algebraic integers are mapped to the lattice $\Lambda_F$, which will be the underlying structure of our codes. Optimal rotations of $\mathbb{Z}^n$ based on ideal lattices arising from canonical embeddings can be found in \cite{Viterbo_tables}. We shall use these algebraic rotations to rotate our coset codes for the simulations, to guarantee a large minimum product distance and hence good performance for Bob. 
\end{remark}

Let $D$ be a positive, square-free integer and and let $F=\mathbb{Q}({\sqrt{D}})$. The discriminant $\Delta$ of $F$ is equal to $ 4 D$ or $D$ depending on whether $D \equiv 2, 3 \pmod 4$ or $D \equiv 1 \pmod 4$ respectively.
The canonical embedding  $\sigma$ of $F$ is determined by 
$\sigma_i: F \longrightarrow \mathbb{R}$ for $i=1,2$ where $\sigma_1(\sqrt{D}) = \sqrt{D}$ and  $\sigma_2(\sqrt{D}) = -\sqrt{D}$ as in Definition \ref{embb}.  
Each element $x + y \sqrt{D} \in F$ with $x, y \in \mathbb{Q}$ is mapped to the vector $(\sigma_1(x + y \sqrt{D}, \sigma_2(x + y \sqrt{D}))=(x + y \sqrt{D},x - y \sqrt{D}) \in \mathbb{R}^2$ via $\sigma$. 
Assume that $I$ has a $\mathbb{Z}$-basis $\{\alpha, \beta\}$. The vectors $\{b_1=(\sigma_1(\alpha), \sigma_2(\alpha)), b_2=(\sigma_1(\beta), \sigma_2(\beta))\}$ form a $\mathbb{Z}$-basis of $\Lambda_I$. In other words, the lattice $\Lambda_I$ can be represented as 
\begin{equation}
\Lambda_I = \begin{bmatrix}
\sigma_1(\alpha) && \sigma_1(\beta)\\
\sigma_2(\alpha)  && \sigma_2(\beta)

\end{bmatrix} \mathbb{Z}^2.
\end{equation}

The following new results are obtained. The proofs are omitted due to lack of space and will appear in an extended version of this paper \cite{HaOliver_journal}. 

\begin{proposition}\label{largenorm}
	If $\Lambda_I$ is a WR ideal lattice in $O_F$, then $N(I)\ge\frac{ \sqrt{ 3 \Delta}}{4}$.
\end{proposition}

In \cite{FukIdeal2},  the authors proved that for imaginary quadratic field $F =\mathbb{Q}(\sqrt{-D})$, the ring of integers contains principal WR ideals if and only if  $D = 1, 3$. For real quadratic fields, we obtain the following result. 

\begin{proposition}\label{Q2}
	There are infinitely many real quadratic fields $F =\mathbb{Q}(\sqrt{D})$ with a positive square-free integer $D \equiv 3 \pmod 4$, respectively $D \equiv 1 \pmod 4$, such that $O_F$ contains principal WR ideals.
\end{proposition}

\begin{remark}
	Determining whether there exists an integer $D \equiv 2 \pmod 4$ such that the quadratic field $\mathbb{Q}({\sqrt{D}})$ has WR principal ideals is still an open problem. A computational search based on Lemma \ref{hermite} showed that $\mathbb{Q}({\sqrt{D}})$ does not contain any WR principal ideals of index at most $2 D$ for any even integer $D \le 10^3$.
\end{remark}


The table below illustrates some examples of WR principal  ideals $\Lambda_I$ of real quadratic fields $F=\mathbb{Q}(\sqrt{D})$ for some positive, square-free integer $D$ given in the first column. The second column contains a generator of principal ideals $I$. The index of the WR lattice $\Lambda_I$ is shown in the last column.

\begin{table}[h]
	\begin{center}
		\label{table1}
		
		\begin{tabular}{cccc}
			\toprule
			\parbox[t]{2mm}{\multirow{6}{*}{\rotatebox[origin=c]{90}{$D \equiv 3\pmod 4 \;\;\;$}} }   &  D & A generator of $I$  & Index\\[0.5ex]  
			\cmidrule(r){2-4}
			&  3 & $3+ \sqrt{3}$ & 6\\ 
			& 15 & $5+ \sqrt{15}$ & 10\\  
			& 35 & $7+ \sqrt{35}$  & 14\\                 
			& 143 & $13+ \sqrt{143}$  & 26\\                                      
			& 195 & $15+ \sqrt{195}$  & 30\\
			
			\hline
			
			\parbox[t]{2mm}{\multirow{6}{*}{\rotatebox[origin=c]{90}{$D \equiv 1\pmod 4 $}} }      &   21 & $\frac{7}{2} -\frac{\sqrt{21}}{2} $  & 7\\ 
			&  77 & $\frac{11}{2} -\frac{\sqrt{77}}{2}$  & 11\\  
			& 165 & $\frac{15}{2} -\frac{\sqrt{165}}{2}$  & 15\\                 
			&221 & $\frac{17}{2} -\frac{\sqrt{221}}{2}$  & 17 \\                                              
			&285 & $\frac{19}{2} -\frac{\sqrt{285}}{2}$  & 19\\
		\end{tabular}
		
	\end{center}
	\caption{Some WR principal  ideals in $\mathbb{Q}(\sqrt{D})$}
\end{table}
%
%
%
%
%
%
%
\section{Simulation Results}
\subsection{Probabilistic Lattice Search}
It is possible to find all WR sublattices of a given lattice and a given index by searching through all possible combinations of vectors of suitable lengths, as described by Lemma \ref{hermite}. Fortunately, WR lattices are common enough that a probabilistic algorithm often suffices.

Most lattices in the simulations were found after only minutes of randomly testing combinations of integer vectors of same length for linear independence and then using the LLL-algorithm to determine $\lambda_1$ for the lattice they generate.

\subsection{Well-Rounded Ideal Lattices}
We simulated the $\ECDP$ for two WR principal ideals and a sublattice of $\ZZ^2$, all having index $216$. More details are given in the table below.
\begin{table}[h!]
	\centering
	\begin{tabular}{c c c c c c}
		\toprule
		Lattice & Field & Ideal & $\lambda_1$ & $R_i$ & $R_c$\\
		\midrule
		$\Lambda_{I_1}$ & $\QQ[\sqrt{3}]$ & $(18+6\sqrt{3})$ & $249.42$ & 3.87744 & 1.12326 \\
		$\Lambda_{I_2}$ & $\QQ[\sqrt{15}]$ & $(18+6\sqrt{15})$ &  $223.08$ & 3.87744 & 1.13166 \\
		$\Lambda_{3}\subset \mathbb{Z}^2$ & - & - & $234$ & 3.87744 & 1.12185 \\
	\end{tabular}
	\caption{Parameters of Lattices after normalization}
\end{table}

\begin{figure}[h]
	\centering
	\includegraphics[scale=0.38]{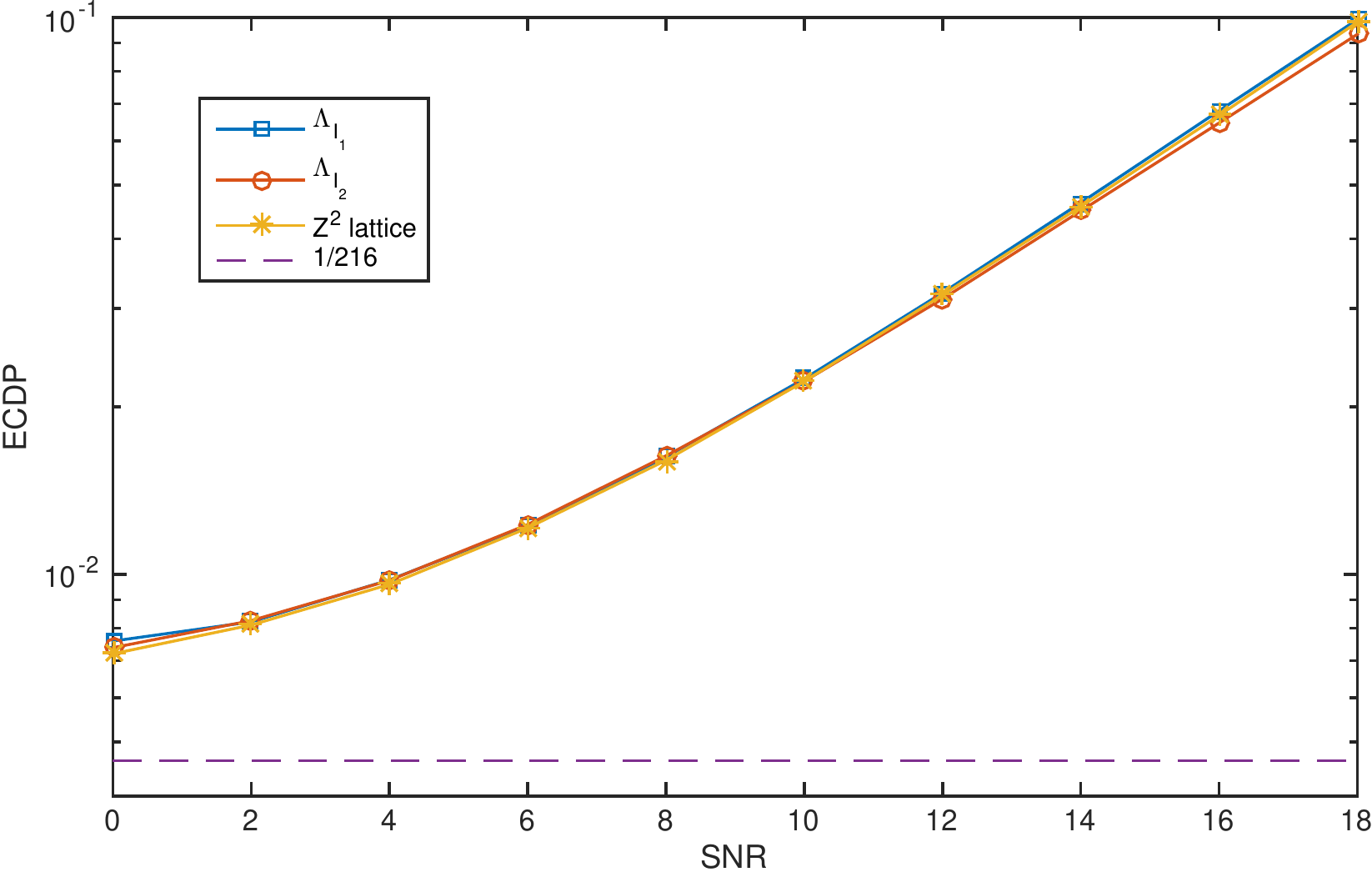}
	\caption{$\ECDP$ for WR ideal lattices and a sublattice of $\mathbb{Z}^2$ with   index $6$}
	\label{Ideal}
\end{figure}

A generator matrix for $\Lambda_3$ is given by $\left(\begin{smallmatrix}
3 & 15 \\ 15 & 3
\end{smallmatrix}\right)$.
In Fig. \ref{Ideal}, the simulation results for the $\ECDP$ are compared for these three lattices and it can be seen that they perform very similarly.
We will therefore look at some higher dimensional examples that allow for significant differences in $\lambda_1$.

\subsection{$\ZZ^4$ sublattices}
In this section, we show simulation results for three different sublattices of $\ZZ^4$,\\

\scalebox{0.76}{$\arraycolsep=2.6pt
	\Lambda_1=\begin{pmatrix}
	16 &0 &0 &0\\
	0&4&0&0\\
	0&0& 2&0\\
	0&0&0& 2
	\end{pmatrix} \ZZ^4, \,\,
	\Lambda_2=\begin{pmatrix}
	4 &0 &0 &0\\
	0&4&0&0\\
	0&0& 4&0\\
	0&0&0&4
	\end{pmatrix} \ZZ^4, \,\,
	\Lambda_3=\begin{pmatrix}
	-2 &-3 &4 &-1\\
	0&-1&0&3\\
	0&-3& -2&-3\\
	-4&-1&0&-1
	\end{pmatrix} \ZZ^4
	$}
\vspace{2pt}
\\
with parameters as shown in Table \ref{Z4params}.

\begin{figure}[h]
	\centering
	\includegraphics[scale=0.50]{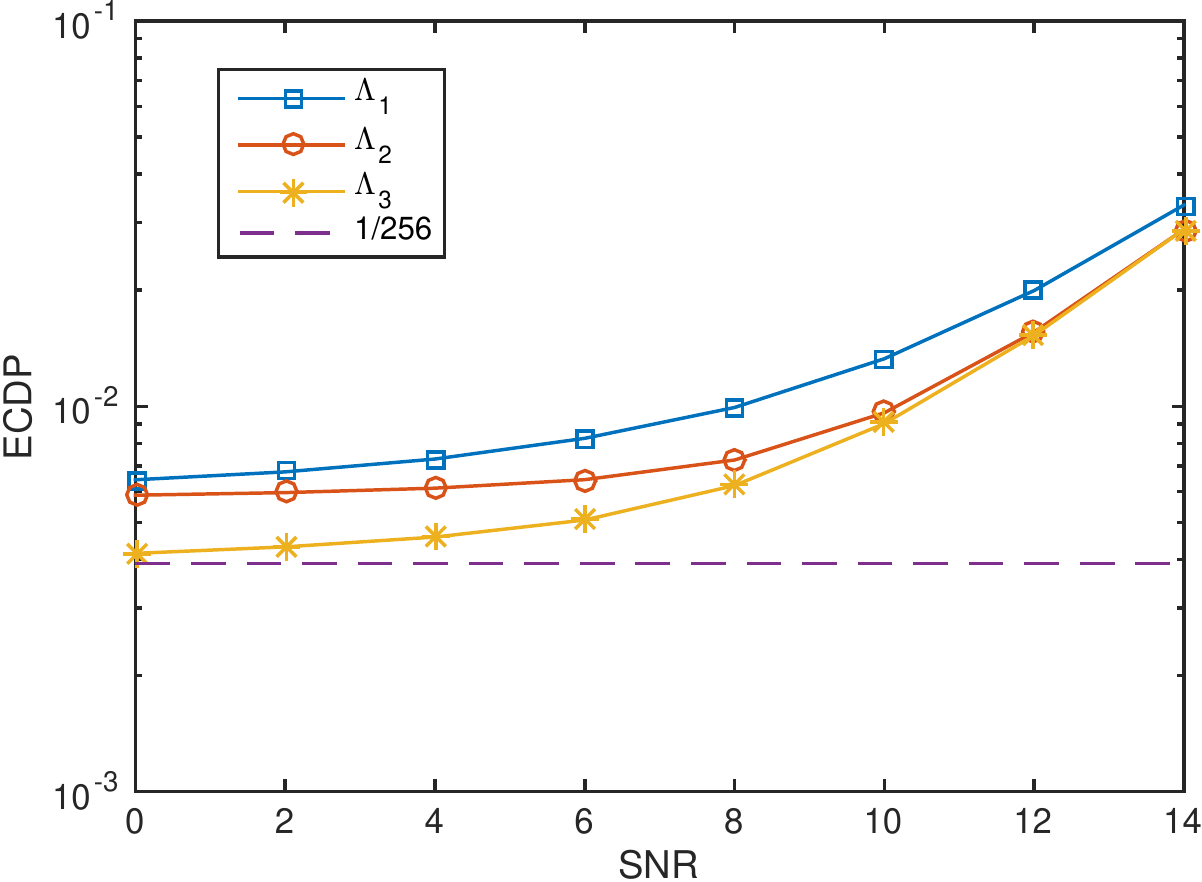}
	\caption{Simulation of $\ECDP$ for three different $\ZZ^4$ sublattices}
	\label{Z4}
\end{figure}

These lattices are chosen because they provide a good sample of different types of lattices with the same information rate and showcase the importance of the $\lambda_1$ parameter.
In Fig. \ref{Z4}, the simulation results for the correct decoding probability $\ECDP$ are shown. It can clearly be seen, that the lattice with larger $\lambda_1$ outperforms the other two and it is the only one that comes close to reaching the theoretical lower bound of $\frac{1}{[\LB : \LE]}$ and hence negligible mutual information. Similar phenomenon can be observed for higher dimensional examples.

\begin{table}[h]
	\centering
	\begin{tabular}{cccccc}
		\toprule
		Lattice	&	$\lambda_1$ & WR & index & $R_i$ & $R_c$ \\
		\midrule
		$\Lambda_1$ &$4$ & no & $256$& $2$ & $2$\\
		$\Lambda_2$ &$16$ & yes & $256$& $2$ & $2$\\
		$\Lambda_3$ &$20$ & yes & $256$& $2$ & $2$\\
		&&&&&
	\end{tabular}
	\caption{Parameters for three different sublattices of $\ZZ^4$}
	\label{Z4params}
\end{table}
\vspace{-0.8cm}
\phantom{x}
\subsection{Cross-Packing}
In a recent paper \cite{AL} a new construction was suggested that aims at minimizing the error probability for Rician channels. This channel model is a generalization of the Gaussian and the Rayleigh channel models and features an additional parameter $K$. For $K=0$ it simplifies to the Rayleigh channel observed here. For small $K$ they observe that the contour lines of the error probability are cross shaped and abstract a cross distance for integer vectors for which they design lattices of minimum cross distance $2t+1$ for all dimensions $n \geq 2$. We compare their cross packing lattice $B_4$ with $t=6$ with a WR sublattice $\Lambda$ of $\ZZ^4$ of the same index $302$ (we use a LLL-reduced basis of $B_4$ here): \\

\scalebox{0.8}{$\qquad \arraycolsep=1.8pt B_4=\begin{pmatrix}
	-1& 1& 2& 2\\
	-1& 0& 2&-5\\
	1&-2& 5&-1\\
	-1&-5& 1& 2
	\end{pmatrix} \ZZ^4, \;
	\Lambda =\begin{pmatrix}
	1& 1& 3&-2\\
	-4& 1& 0&-4\\
	-1&-2& 3& 1\\
	2&-4& 2&-1
	\end{pmatrix}\ZZ^4 $} \\

\begin{figure}[h!]
	\centering
	\includegraphics[scale=0.5]{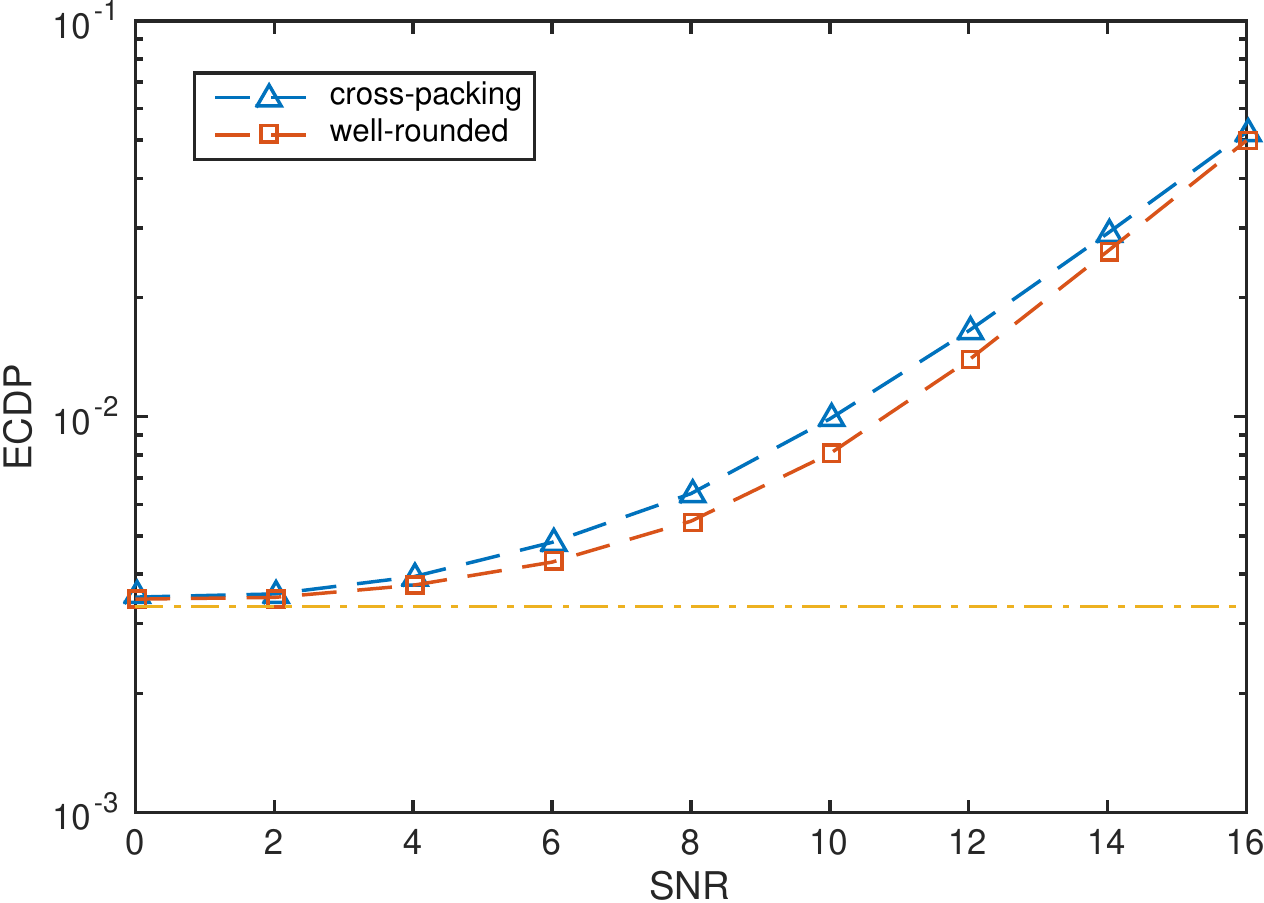}
	\caption{Comparison of a Cross-Packing Lattice with a WR lattice}
	\label{cross}
\end{figure}
Fig. \ref{cross} shows that the WR lattice outperforms the cross packing. We also point out that the cross-packing construction is only available for indices $t^3+2t^2+2t+2$ with $t \in \NN$.




\section*{Acknowledgment}

This work was partially supported by the Academy of Finland
grants \#276031, \#282938, \#283262, and \#303819, and by a grant
from Magnus Ehrnrooth Foundation, Finland. The support
from the European Science Foundation under the ESF COST
Action IC1104 is also gratefully acknowledged.



%

\bibliographystyle{IEEEtran}
\bibliography{IEEEabrv,References}

\begin{thebibliography}{10}
\providecommand{\url}[1]{#1}
\csname url@samestyle\endcsname
\providecommand{\newblock}{\relax}
\providecommand{\bibinfo}[2]{#2}
\providecommand{\BIBentrySTDinterwordspacing}{\spaceskip=0pt\relax}
\providecommand{\BIBentryALTinterwordstretchfactor}{4}
\providecommand{\BIBentryALTinterwordspacing}{\spaceskip=\fontdimen2\font plus
\BIBentryALTinterwordstretchfactor\fontdimen3\font minus
  \fontdimen4\font\relax}
\providecommand{\BIBforeignlanguage}[2]{{%
\expandafter\ifx\csname l@#1\endcsname\relax
\typeout{** WARNING: IEEEtran.bst: No hyphenation pattern has been}%
\typeout{** loaded for the language `#1'. Using the pattern for}%
\typeout{** the default language instead.}%
\else
\language=\csname l@#1\endcsname
\fi
#2}}
\providecommand{\BIBdecl}{\relax}
\BIBdecl

\bibitem{BelfioreOggierRayleigh}
J.~C. Belfiore and F.~Oggier, ``Lattice code design for the rayleigh fading
  wiretap channel,'' in \emph{IEEE Int. Comm. Workshop (ICC)}, June 2011, pp.
  1--5.

\bibitem{Mirghasemi}
H.~Mirghasemi and J.~Belfiore, ``Lattice code design criterion for {MIMO}
  wiretap channels,'' in \emph{2015 {IEEE} Information Theory Workshop - Fall
  (ITW)}, 2015, pp. 277--281.

\bibitem{ling}
C.~Ling, L.~Luzzi, J.-C. Belfiore, and D.~Stehl{\'e}, ``Semantically secure
  lattice codes for the gaussian wiretap channel,'' \emph{IEEE Transactions on
  Information Theory}, vol.~60, no.~10, pp. 6399--6416, 2014.

\bibitem{luzzi}
L.~Luzzi, C.~Ling, and R.~Vehkalahti, ``Almost universal codes for fading
  wiretap channels,'' \emph{arXiv preprint arXiv:1601.02391}, 2016.

\bibitem{AlexFlat}
\BIBentryALTinterwordspacing
A.~Karrila, A.~Barreal, D.~A. Karpuk, and C.~Hollanti, ``Information bounds and
  flatness factor approximation for fading wiretap channels,'' 2016. [Online].
  Available: \url{http://arxiv.org/abs/1606.06099}
\BIBentrySTDinterwordspacing

\bibitem{AlexSkew}
A.~Karrila and C.~Hollanti, ``A comparison of skewed and orthogonal lattices in
  gaussian wiretap channels,'' in \emph{{IEEE} Inf. Theory Workshop, {ITW}
  2015, Jerusalem, Israel, April 26 - May 1, 2015}, 2015, pp. 1--5.

\bibitem{Wyner}
A.~D. Wyner, ``The wire-tap channel,'' \emph{The Bell System Technical
  Journal}, vol.~54, no.~8, pp. 1355--1387, Oct 1975.

\bibitem{InfoSec}
Y.~Liang, H.~V. Poor, and S.~S. (Shitz), ``Information theoretic security,''
  \emph{Foundations and Trends® in Communications and Information Theory},
  vol.~5, no. 4–5, pp. 355--580, 2008.

\bibitem{Viterbo_green}
F.~Oggier and E.~Viterbo, ``Algebraic number theory and code design for
  rayleigh fading channels,'' \emph{Foundations and Trends® in Comm. and Inf.
  Theory}, vol.~1, no.~3, pp. 333--415, 2004.

\bibitem{OggierCoset}
F.~Oggier, P.~Sole, and J.~C. Belfiore, ``Lattice codes for the wiretap
  gaussian channel: Construction and analysis,'' \emph{IEEE Transactions on
  Information Theory}, vol.~PP, no.~99, pp. 1--1, 2015.

\bibitem{Wyner2}
L.~H. Ozarow and A.~D. Wyner, \emph{Advances in Cryptology: Proceedings of
  EUROCRYPT 84}.\hskip 1em plus 0.5em minus 0.4em\relax Springer Berlin
  Heidelberg, 1985, ch. Wire-Tap Channel II, pp. 33--50.

\bibitem{Karpuk2014}
D.~A. Karpuk, A.~Ernvall{-}Hyt{\"{o}}nen, C.~Hollanti, and E.~Viterbo,
  ``Probability estimates for fading and wiretap channels from ideal class zeta
  functions,'' \emph{Adv. in Math. of Comm.}, 2015, arxiv:1412.6946.

\bibitem{Roope}
R.~Vehkalahti, H.-F. Lu, and L.~Luzzi, ``Inverse determinant sums and
  connections between fading channel information theory and algebra,''
  \emph{{IEEE} Trans. Inf. Theory}, vol.~59, no.~9, pp. 6060--6082, 2013.

\bibitem{ithgain}
R.~Vehkalahti and C.~Hollanti, ``Reducing complexity with less than minimum
  delay space-time lattice codes,'' in \emph{Information Theory Workshop (ITW),
  2011 IEEE}, Oct 2011, pp. 130--134.

\bibitem{DaveRotate}
D.~A. Karpuk and C.~Hollanti, ``Locally diverse constellations from the special
  orthogonal group,'' \emph{IEEE Transactions on Wireless Communications},
  vol.~15, no.~6, pp. 4426--4437, June 2016.

\bibitem{PhongNguyen}
P.~Q. Nguyen, \emph{The LLL Algorithm: Survey and Applications}.\hskip 1em plus
  0.5em minus 0.4em\relax Springer Berlin Heidelberg, 2010, ch. Hermite's
  Constant and Lattice Algorithms, pp. 19--69.

\bibitem{McMullenMinkowski}
C.~T. McMullen, ``Minkowski's conjecture, well-rounded lattices and topological
  dimension,'' \emph{J. Amer. Math. Soc.}, vol.~18, no.~03, pp. 711--735, jul
  2005.

\bibitem{Blichfeldt}
H.~F. Blichfeldt, ``The minimum value of quadratic forms, and the closest
  packing of spheres,'' \emph{Math. Ann.}, vol. 101, no.~1, pp. 605--608, 1929.

\bibitem{Martinet-Jacques}
J.~Martinet, \emph{Perfect lattices in {E}uclidean spaces}, ser. Grundlehren
  der Mathematischen Wissenschaften.\hskip 1em plus 0.5em minus 0.4em\relax
  Springer-Verlag, Berlin, 2003, vol. 327.

\bibitem{FukIdeal2}
L.~Fukshansky, G.~Henshaw, P.~Liao, M.~Prince, X.~Sun, and S.~Whitehead, ``On
  well-rounded ideal lattices ii,'' \emph{Int. J. Number Theory}, vol.~09,
  no.~01, pp. 139--154, 2013.

\bibitem{FukIdeal}
L.~Fukshansky and K.~Petersen, ``On well-rounded ideal lattices,'' \emph{Int.
  J. Number Theory}, vol.~8, no.~1, pp. 189--206, 2012.

\bibitem{Viterbo_tables}
E.~Viterbo, ``Full-diversity rotations,'' webpage:
  http://www.ecse.monash.edu.au/staff/eviterbo/rotations/rotations.html.

\bibitem{HaOliver_journal}
O.~Gnilke, H.~T.~N. Tran, A.~Karrila, and C.~Hollanti, ``Well-rounded lattices
  for reliability and security in {SISO} and {MIMO} rayleigh fading channels,''
  in preparation.

\bibitem{AL}
A.~Sakzad, A.~L. Trautmann, and E.~Viterbo, ``Cross-packing lattices for the
  rician fading channel,'' in \emph{IEEE Inf. Theory Workshop (ITW)}, April
  2015, pp. 1--5.

\end{thebibliography}

\end{document}